# Influence of growth parameters on structural properties and bandgap of InN epilayers deposited in a showerhead MOVPE system

Abdul Kadir[1,*], Tapas Ganguli[2], Ravi Kumar[2], M.R. Gokhale[1], A.P. Shah[1], B.M. Arora[1] and Arnab Bhattacharya[1]

[1]*Tata Institute of Fundamental Research, Homi Bhabha Road, Mumbai 400005, India*
[2]*Raja Ramanna Center for Advanced Technology, Indore 425013, India*
[*]*Email: abdulkadir@tifr.res.in*

Date: 30[th] April 2007

*Abstract:* From a detailed analysis of InN epilayers deposited in a close-coupled showerhead metalorganic vapour phase epitaxy (MOVPE) system under various conditions we investigate the effect of growth parameters on the lattice constants of the InN layer. The layers are under significant internal hydrostatic stress which influences the optical properties. Samples typically fall into two broad categories of stress, with resultant luminescence emission around 0.8eV and 1.1eV. We can correlate the internal stress in the layer and the value of the optical absorption edge, and the PL emission wavelength.

*Introduction*

While there has been a significant improvement in the quality of MOVPE grown InN epilayers, the material typically has a large background concentration and there is still uncertainty on the cause of variations in bandgap reported by different groups. A large body of recent work suggests that the bandgap of InN is of the order of 0.7eV-0.8eV [1,2]. There is, however, uncertainty on the cause of variations in InN bandgap reported by different groups which has been attributed to a Moss-Burstein shift[3,4], oxygen alloying[4], presence of metallic In and Mie scattering from such In droplets[5], presence of trapping levels[6] and quantum size effects[7] etc. Compared to the large volume of literature on the bandgap controversy, there is relatively little published data that discusses any systematic dependence of the measured bandgap of InN to the lattice parameter(s). The scatter in the reported values of the "*c*" and "*a*" lattice constants of InN is quite large and has mostly been attributed to the inbuilt strain (biaxial) in the InN layers due to the effects of the substrate/ buffer[8]. There are only few reports on the presence of hydrostatic strain in InN films [9]. We have earlier reported a detailed study of the growth parameter space for InN in a close-coupled showerhead MOVPE system[10]. As a result of this study we have access to over 40 InN samples grown under different MOVPE conditions. In this work we investigate the effect of MOVPE growth parameters on variation in lattice constants which in turn influence the strain and optical properties of InN epilayers.

*Experimental*

All the InN epilayers studied were synthesized by MOVPE on 2" *c*-plane sapphire substrates in a 3x2" close coupled showerhead reactor (Thomas Swan) using trimethylindium (TMIn) and ammonia (NH$_3$) precursors with N$_2$ carrier gas. Growth parameters such as temperature[11], T$_g$, (500–570°C), reactor pressure, R$_p$, (200–700 Torr), V/III ratio (5000–37000), and TMIn flux (6–12μmol/min.) were varied for different layers[10]. The showerhead-susceptor spacing was 11mm, typical for nitride growth. Before the growth of the InN layer, a 1μm thick undoped GaN buffer layer was grown at 1040°C using a standard two-step process. The temperature was then ramped down to the InN growth temperature, and the carrier gases switched over to N$_2$. The total flow through the showerhead was 12 slpm. In addition, for the samples grown directly on sapphire without the GaN buffer, a two-step growth process using a low-temperature buffer layer was used [12].

The InN films were structurally characterized by high resolution X-ray diffraction (HRXRD) on a PANalytical X-pert MRD system with a Hybrid 4-bounce monochromator. This system was used for the determination of the lattice parameters based on the method described in Ref. 13, which corrects for the errors due to the centering of the samples on the goniometer. Other corrections e.g. due to refractive index, Lorentz-polarization and absorption together lead to an estimated lattice parameter inaccuracy of ~1x10$^{-5}$Å, and were neglected as the differences in the lattice parameters observed in this work are >1%. The strain in the InN layers is evaluated using the measured lattice constants and the unstrained lattice parameter value of 5.7064Å and 3.5376Å for "*c*" and "*a*" respectively[14]. The stress tensor components in the samples are then evaluated from the values of the compliance tensor for InN [15,16].

| Sample | V/III | $T_g$ (°C) | Buffer | c-axis (Å) | a-axis (Å) | $E_{xx}$ | $E_{zz}$ | Absorp. edge(eV) | PL peak (eV) |
|---|---|---|---|---|---|---|---|---|---|
| 05027 | 5095 | 530 | GaN | 5.5131 | 3.4078 | -0.03669 | -0.03387 | 1.37 | No PL |
| 05028 | 10190 | 530 | GaN | 5.5134 | 3.4346 | -0.02912 | -0.03382 | 1.41 | 1.09 |
| 05040 | 18700 | 530 | GaN | 5.6183 | 3.483 | -0.01543 | -0.01544 | 1.02 | 0.84 |
| 06037 | 25000 | 530 | GaN | 5.5848 | 3.464 | -0.02081 | -0.02131 | 1.21 | 0.94 |
| 06036 | 37000 | 530 | GaN | 5.5937 | 3.4667 | -0.02004 | -0.01975 | 1.08 | 0.85 |
| 05039 | 10190 | 570 | GaN | 5.6287 | 3.4885 | -0.01388 | -0.01361 | 1.0 | 0.82 |
| 05030 | 10190 | 550 | GaN | 5.5312 | 3.419 | -0.03353 | -0.0307 | 1.31 | 1.08 |
| 05032 | 10190 | 530 | none | 5.5535 | 3.429 | -0.0307 | -0.02679 | 1.14 | 1.08 |
| 06033 | 18700 | 530 | none | 5.6150 | 3.494 | -0.01232 | -0.01602 | 1.04 | 0.84 |
| 06004 | 18700 | 530 | GaN | 5.6445 | 3.498 | -0.01119 | -0.01085 | 0.94 | 0.81 |
| 05033 | 10190 | 510 | none | 5.5202 | 3.4217 | -0.03276 | -0.03264 | 1.40 | No PL |
| 05029 | 10190 | 510 | GaN | 5.5201 | 3.419 | -0.03353 | -0.03265 | 1.34 | No PL |

*Table 1: Details of growth parameters, lattice constants, and optical properties of InN samples discussed in Fig.2. All samples were grown at 300 Torr, and are 0.2μm thick apart from sample 06004 which is 0.45μm thick. Samples with "none" under buffer were grown directly on sapphire. $E_{xx}$ and $E_{zz}$ represent in-plane and out-of-plane strains respectively.*

Room/low temperature photoluminescence (PL) measured using a 0.67m monochromator with $Ar^+$ ion laser excitation, and absorption measurements (on back-side polished samples) on a Cary 5000 UV/VIS/NIR spectrophotometer were used to evaluate the bandgap. Hall measurements in the Van-der-Pauw geometry were used to determine the carrier concentration and the carrier mobility.

*Results and Discussion*

Fig. 1a shows ω-2θ x-ray scans for 4 representative 0.2μm thick InN layers deposited at 300 Torr and 530°C on sapphire with a GaN buffer, at varying V/III ratios. The inset shows details of the InN (0002) peak which varies for different layers. Broadly, the 2θ positions fall in two categories: (a) samples having a 2θ value ~32.4° and (b) samples having a 2θ value ~32°. This depends critically on the growth conditions, for example, larger 2θ values (smaller lattice constant) are obtained from layers deposited at V/III ratios ~10,000 or less, while ones deposited at higher V/III ratios have smaller 2θ values, corresponding to slightly larger values of lattice constants. The InN layers in category (b) also show signatures of free In: In(101) at 2θ=32.94°. Identical HRXRD patterns from InN layers before and after dipping in dilute HCl confirm that the free In is present at the grain boundaries and not the InN sample surface. The corresponding absorption curves near the band edge (Fig. 1(b)) and the room temperature photoluminescence spectra (Fig. 1(c)) also show that there seem to be two broad categories of samples. The absorption spectra show band edges around 1.0-1.1 eV, and around 1.3-1.4 eV for the samples having larger and smaller lattice constants respectively. This trend is also seen in the PL spectra, samples with larger lattice constants have PL peaks ~0.82 eV while samples with smaller lattice constant have (weaker) PL with emission centered at ~1.1eV. The latter samples are grown at a relatively low V/III ratio, not optimal for high quality InN, and the weaker luminescence possibly results from a higher concentration of defects.

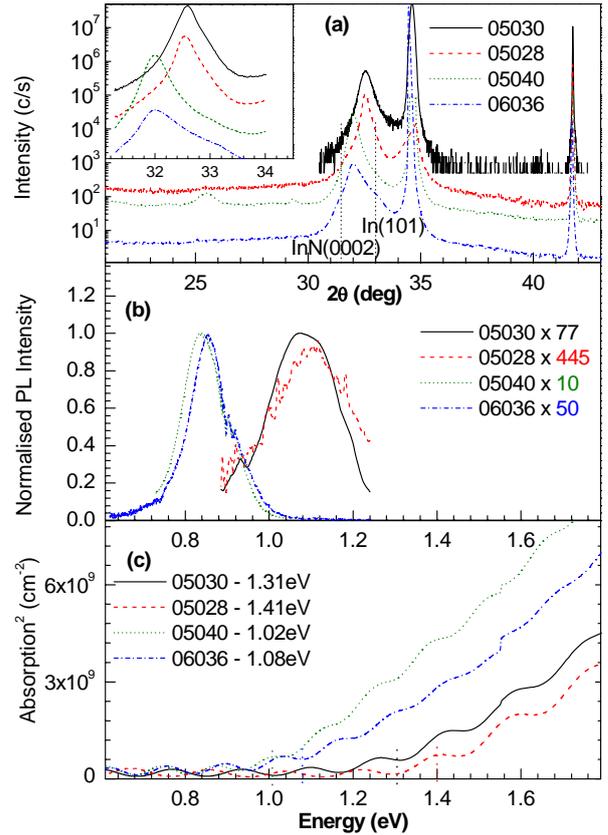

*Fig. 1: (a) X-ray ω-2θ scans (b) normalized low-temp. PL spectra (c) absorption spectra of 0.2μm-thick InN layers. Inset shows structure near InN (0002) peak. See text for details.*

The lattice parameters (*c* and *a* values) for the various layers were evaluated using data from different *(hkl)* reflections. Thereafter the strain is calculated and the corresponding stress evaluated. The samples fall into two broad categories of lattice parameters and have compressive stress (a) ~ 4-5 GPa and (b) ~12-14 GPa. The presence of such a large stress in the InN layer should shift the band edge in addition to the effects due to the large carrier concentration. To further analyze the nature of stress and to eliminate other possible causes for the shift in lattice constant we have examined a range of samples grown under different conditions (Table 1). This data is plotted in Fig. 2, which shows the relative shift in lattice constant from the "unstrained" literature value[14] along the *c*- and *a*-axes, for 11 samples. The diagonal lines show the expected changes in lattice constants for purely hydrostatic or purely biaxial strains. Points shown as open circles are InN epilayers grown directly on sapphire without a GaN buffer. For comparison, the values for MBE-grown nearly unstrained sample from two groups (Refs. 8,14) are also shown. From the scatter in the points it can be clearly seen that most samples lie close to the line of pure hydrostatic strain, with a relatively small biaxial component. While there is a small dispersion, many of the samples seem to fall into two categories, of about 1% and 3% strain respectively, consistent with the picture seen in the representative samples discussed in Fig. 1. It should be pointed out that InN layers grown directly on sapphire without a GaN buffer also fall into the two categories. This rules out InGaN alloy formation due to intermixing of gallium from the GaN buffer as the cause of any shift in InN 2θ peak position. Hence, the lattice constant is determined primarily by the growth parameters for the InN layer, and not influenced by the buffer layer. Further, the presence of $In_2O_3$ from oxygen contamination is highly unlikely in a modern leak-tight MOVPE system that grows high-quality AlGaN layers.

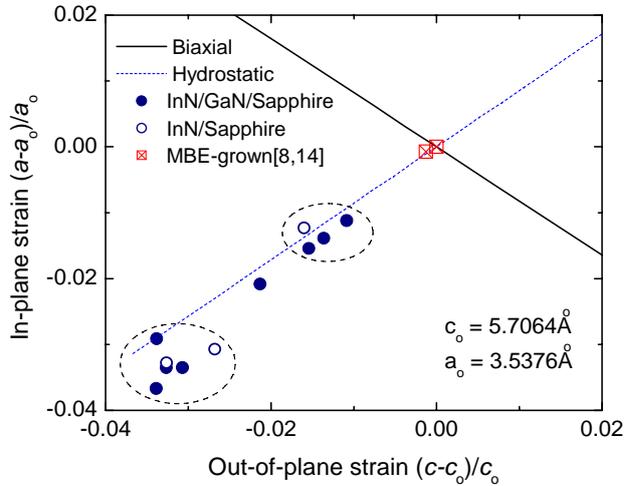

*Fig. 2: Relative shift in a- and c-axis lattice parameters of InN epilayers grown under various conditions. Lines indicate shifts for pure biaxial/ hydrostatic strains.*

From Hall measurements, the values of the background concentration for the films whose data is plotted in Fig. 2 varies between $1.6 \times 10^{19} cm^{-3}$ and $4.3 \times 10^{19} cm^{-3}$. Assuming $E_g(0)$ of InN to be 0.7eV, and using the non-parabolic model [3] the shift in band edge due to the Burstein-Moss shift for such a change in carrier concentration is estimated to be between 0.14eV and 0.28eV. This shift alone cannot explain the observed variation in the value of the absorption edge from ~1.0 to 1.4eV.

The resultant stress from the presence of large (1%-3%) strains in the InN layer should shift the band edge in addition to the effects due to the large carrier concentration. To estimate the influence of stress on the electronic band structure, we have calculated the InN band edge shift for our samples using the Bir-Pikus Hamiltonian and evaluated transition energies involving the conduction band and the topmost strain modified valence band[17]. We find that the internal hydrostatic stress strongly influences the shift in bandedge, and the trend seen in the stress dependence of bandgap are similar to that reported on studies[18] of the effect of *externally* applied hydrostatic pressure on the band edge of InN.

The knowledge of the MOVPE growth conditions that lead to relatively high- or low-stresses in the InN layer allow us to make a reasonable conjecture on the probably causes underlying this. We believe that the stress is determined primarily by the nitrogen vacancies in the layer[13]. A decrease in lattice parameter, for both *c*- and *a*-axis, due to nitrogen vacancies has been reported for GaN as well [19]. At high V/III ratios (i.e. high ammonia flow), or at higher growth temperature (better ammonia cracking) the amount of available nitrogen species at the growth surface is higher, thus reducing the number of nitrogen vacancies, and hence the deformation of the unit cell, and the stress in the layer. We also observe traces of free Indium in samples grown at high V/III ratio. This is not surprising, as more available nitrogen concomitantly also implies the presence of more atomic hydrogen from $NH_3$ decomposition. It is well known that atomic hydrogen impedes the incorporation of Indium in InGaN. The presence of more H in the reactor leads to In atoms beings expulsed from the InN lattice resulting in the presence of free In.

In conclusion, by carefully measuring the lattice constants of MOVPE grown InN epilayers we see a correlation between the internal hydrostatic stress in the layer and the value of the optical absorption edge, and the PL emission wavelength. We find that our MOVPE grown samples typically fall into two broad categories of stress, with resultant PL emission around 0.8eV and 1.1eV, depending on the amount of available nitrogen. This stress-related shift may be important in determining the optical properties of InN layers.